\documentclass[aps,prl,twocolumn,groupedaddress]{revtex4}
\usepackage{graphicx,bm,amssymb}

\begin{document}
\title{Spin relaxation and decoherence of holes in quantum dots}
\author{Denis V. Bulaev and Daniel Loss}
\affiliation{Department of Physics and Astronomy, University of Basel, Klingelbergstrasse 82, CH-4056 Basel,
Switzerland}
\date{\today}

\begin{abstract} 
We investigate heavy-hole spin relaxation and decoherence in quantum dots in perpendicular magnetic fields. We show that at low temperatures the spin decoherence time is two times longer than the spin relaxation time. We find that the spin relaxation time for heavy holes can be comparable to or even longer than that for electrons in strongly two-dimensional quantum dots. We discuss the difference in the magnetic-field dependence of the spin relaxation rate due to Rashba or Dresselhaus spin-orbit coupling for systems with positive (i.e., GaAs quantum dots) or negative (i.e., InAs quantum dots) $g$-factor.
\end{abstract}


\maketitle

Spin physics has become one of the most rapidly
developing branches of condensed matter physics. Spin physics is very important, not only from a fundamental point of view, but also for the fabrication of novel electronic devices, for the experimental realization of quantum computation, and for the development of spin electronics (spintronics) \cite{ALS}. Quantum dots (QDs) are most attractive candidates for these applications because of their
reduced dimensionality, leading to long-lived spin states and allowing single spin manipulation \cite{LV}.

Recent experiments \cite{RBMCh,E2004,K2004} show that electrons in QDs have a long spin relaxation time (up to $20\:$ms \cite{K2004}) and it is now possible to prepare a single electron spin state with a well-defined orientation, read the spin state out, and store the information about the spin orientation for a long time \cite{K2004}. There are two main spin relaxation mechanisms for electron spins in QDs: that due to the electron-phonon interaction \cite{KN,GKL,BL,FA} and that due to the hyperfine interaction with surrounding  nuclear spins \cite{EN,KLG,CL}. Since the valence band has $p$ symmetry, the hyperfine interaction of holes with lattice nuclei is suppressed with respect to that of the conduction band (electrons). This has led to an increased interest in hole spins as carries of long-lived quantum information. It was shown that in thin quantum wells (QWs) the hole spin relaxation is slower than that in the bulk case \cite{SKG,SFS}. Nevertheless, the hole spin relaxation time is several orders of magnitude smaller than that for electrons. This is due to the fact that, in addition to existing spin-orbit (SO) couplings for electrons due to bulk inversion asymmetry (BIA) (Dresselhaus spin-orbit (DSO) coupling \cite{Dresselhaus}) and structure inversion asymmetry (SIA) (the Rashba spin-orbit  (RSO) coupling \cite{BR}) there is strong SO coupling between the heavy-hole (HH) and light-hole (LH) subbands \cite{ER}.

Very recently, investigation of hole spin relaxation in QDs was reported \cite{WRK,LCW}. In these works only one SO mechanism was considered, the SO coupling between HHs and LHs. It was shown that the hole spin relaxation time in QDs is longer than that in QWs but still shorter by several orders of magnitude than that for electrons in QDs. Furthermore, it was found that SO coupling between HHs and LHs is negligible for two-dimensional (2D) QDs if the energy splitting between the HH and LH subbands is much larger than the level spacing in those subbands \cite{LCW}. Up to now this case has not been investigated, though it is very important for the realization of coherent spin states with a spin relaxation time longer than that for electrons. In this case other SO coupling mechanisms (RSO and DSO couplings) become significant.

In this Letter, we study HHs confined to a QD in a perpendicular magnetic field. We consider the three main SO coupling mechanisms: RSO, DSO, and SO coupling between the HH and LH subbands. An effective Hamiltonian for 2D HHs is derived. We study the spin relaxation and decoherence of the system induced by the interaction of HHs with phonons.

From the two-band Kane model, the Hamiltonian for the valence band of III--V semiconductors is given by \cite{OO}
\begin{equation}
\label{eq:Hbulk} H_{bulk}=H_{LK}-\frac{\gamma}{\eta}\mathbf{J}\cdot\bm{\Omega},
\end{equation}
where $H_{LK}$ is the Luttinger -- Kohn Hamiltonian \cite{LK}, $\gamma$ is due to BIA,
$\eta=\Delta_{so}/(E_g+\Delta_{so})$, $\Delta_{so}$ is the split-off gap energy, $E_g$ is the band gap energy,
$\textbf{J}=(J_x,J_y,J_z)$ are $4\times4$ matrices corresponding to spin $3/2$, $\Omega_z=P_z(P_x^2-P_y^2)$, and
$\Omega_x$, $\Omega_y$ are given by cyclic permutations. The last term in Eq.~(\ref{eq:Hbulk}), caused by SO interaction
of the conduction and valence bands, is DSO coupling for the valence band (for the conduction band, the DSO coupling
is given by \cite{Dresselhaus} $\gamma\bm{\sigma}\cdot\bm{\Omega}$, where $\bm{\sigma}=(\sigma_x,\sigma_y,\sigma_z)$ is
the vector of Pauli matrices). The magnetic field induces a Zeeman splitting, which is described by the following
term \cite{Luttinger}: $H_Z=-2\kappa\mu_B\mathbf{B\cdot J}-2q\mu_B\mathbf{B}\cdot\bm{\mathcal{J}}$, where $\kappa$ and
$q$ are the Luttinger parameters \cite{Luttinger} and $\bm{\mathcal{J}}=(J_x^3,J_y^3,J_z^3)$. For 2D asymmetric QWs, due
to SIA along the growth direction, there is an additional SO term, the RSO term, which in the two-band model is given
by \cite{Winkler1,Winkler2} $\alpha_R\mathbf{P\times E\cdot J}$, where $\alpha_R$ is the RSO coupling constant and
$\mathbf{E}$ is an effective electric field along the growth direction.

We consider a $[001]$-grown 2D system. Due to confinement along the growth direction, the valence
band splits into a HH subband with $J_z=\pm3/2$ and a LH subband with $J_z=\pm1/2$ \cite{ER}. If the splitting of HH and LH subbands is
large, we describe the properties of HHs and LHs separately, using only the $2\times2$ submatrices for the
$J_{z}=\pm3/2$ and $J_{z}=\pm1/2$ states, respectively. The HH submatrices have the property that
$\tilde{J}_{x}=\tilde{J}_{y}=0$ and $\tilde{J}_{z}=\frac32\sigma_z$ (see Ref.~\onlinecite{KCPF}). For such a system and
low temperatures only the lowest HH subband is significantly occupied. In this case, we consider HHs
only. In the framework of perturbation theory \cite{OO}, using Eq.~(\ref{eq:Hbulk}) and taking into account the Zeeman
energy, the RSO, and DSO term, the effective Hamiltonian for HHs of a QD with lateral confinement potential $U(x,y)$ is given
by
\begin{equation}
\label{eq:H} H=\frac{1}{2m}(P_x^2+P_y^2)+U(x,y)+H_{so}^{hh}-\frac12g\mu_B\mathbf{B}\cdot\bm{\sigma},
\end{equation}
where $m=m_0/(\gamma_0+\gamma_1)$ is the effective HH mass, $m_0$ is the free electron mass, $\gamma_0$ and $\gamma_1$
are the Lutinger parameters \cite{Luttinger}, $g$ is the effective $g$-factor of HHs, and 
\begin{equation}
\label{eq:Hso} H_{so}^{hh}=i\alpha(\sigma_+P_-^3-\sigma_-P_+^3)+\beta(\sigma_+P_-P_+P_-+\sigma_-P_+P_-P_+)
\end{equation}
is the SO coupling of HHs, which is due to the SO coupling between LH and HH subbands, SIA (the first term), and BIA (the
second term). The first term in Eq.~(\ref{eq:Hso}) is the RSO coupling \cite{Winkler1,Winkler2} and the second term is
the DSO coupling. Here, $\alpha=3\gamma_0\alpha_R\langle E_z\rangle/2m_0\Delta$, $\beta=3\gamma_0\gamma\langle
P_z^2\rangle/2m_0\eta\Delta$, $\sigma_\pm=(\sigma_x\pm i\sigma_y)/2$, $P_\pm=P_x\pm iP_y$, $\langle E_z\rangle$ is the
averaged effective electric field along the growth direction of a QD, and $\Delta$ is the splitting of LH and HH
subbands. The splitting between HH and LH subbands $\Delta\sim d^{-2}$, where $d$ is the QD height.
Comparing Eq.~(\ref{eq:Hso}) with the SO coupling term for electrons, we find that for a QD with the characteristic
lateral size $l$, the ratio $\langle H_{so}^{el}\rangle/\langle H_{so}^{hh}\rangle\sim(l/d)^2$. Thus, for strongly 2D QDs
($l\gg d$), the SO coupling of HHs can be less than that for electrons. 

Without the SO interaction ($\alpha=\beta=0$), the spectrum of a QD Hamiltonian with parabolic lateral confinement
can be found using a canonical transformation of the Hamiltonian \cite{GMSh}. For circularly-symmetric QDs with
oscillator frequency $\omega_0$ ($U(x,y)=m\omega_0^2(x^2+y^2)$) in a perpendicular magnetic field, the energy spectrum
and wave functions of HHs are given by \cite{BL}
$E_{n_1n_2\uparrow(\downarrow)}=\hbar\omega_-(n_1+1/2)+\hbar\omega_+(n_2+1/2)\mp\hbar\omega_Z/2$,
$\left|n_1n_2\uparrow(\downarrow)\right\rangle=\Phi_{n_1}(q_1\sqrt{m\omega_-/\hbar})\Phi_{n_2}(q_2\sqrt{m\omega_+/\hbar})\left%
|\uparrow(\downarrow)\right\rangle$, where $n_1,n_2=0,1,2,\ldots$, $\omega_\pm=\Omega\pm\omega_c/2$ are the hybrid
frequencies, $\Omega=\sqrt{\omega_0^2+\omega_c^2/4}$, $\omega_c=|e|B/mc$ is the cyclotron frequency,
$\omega_Z=g\mu_BB/\hbar$ is the Zeeman frequency, $q_1$ and $q_2$ are new coordinates in the transformed phase space \cite{GMSh},
and $\Phi_n(q)$ are oscillator functions. We only consider low-lying levels. For definiteness, we assume that $B>0$ and
$g>0$, then the ground state is the spin-up state $|00\uparrow\rangle$. At low $B$, the next level is
$E_{00\downarrow}$, which is Zeeman-split from the ground state level, and at high $B$, levels $E_{n0\uparrow}$ are
close to the ground state level (since $\omega_-\to0$ as $B\to\infty$). Therefore, there are crossings of levels $E_{n0\uparrow}$
with $E_{00\downarrow}$ at $\omega_Z=n\omega_-$.

We now take the SO coupling of HHs into account. The spectral problem for $H$ can be solved in the framework of
perturbation theory \cite{BL}. The SO coupling influences the wave functions more than the energy spectrum (since
the energy corrections due to $H_{so}^{hh}$ are only second order). Thus, SO coupling leads to mixing of spin-up and spin-down
states. The RSO and DSO terms differ by symmetry in momentum space \cite{GBG,BL} and hence mix different
states: the state $|00\uparrow\rangle$ mixes with $|03\downarrow\rangle$
($E_{03\downarrow}-E_{00\uparrow}=\hbar(3\omega_++\omega_Z)$) due to RSO and with the states $|01\downarrow\rangle$
($E_{01\downarrow}-E_{00\uparrow}=\hbar(\omega_++\omega_Z)$) and $|12\downarrow\rangle$ due to DSO coupling. In 
turn, $|00\downarrow\rangle$ mixes with $|30\uparrow\rangle$
($E_{30\uparrow}-E_{00\downarrow}=\hbar(3\omega_--\omega_Z)$) due to RSO and with $|10\uparrow\rangle$
($E_{10\uparrow}-E_{00\downarrow}=\hbar(\omega_--\omega_Z)$) and $|21\uparrow\rangle$ due to DSO coupling. Again, we only
consider the case $B>0$, since the physical properties of the system are independent of the sign of $B$ \cite{BL}. In this case, $\omega_+>|\omega_Z|$ and the mixed state levels cross (at $\omega_Z=\omega_-$ and $\omega_Z=3\omega_-$) for positive HH $g$-factor ($\omega_Z>0$) but not for
$g<0$. Therefore, for $g>0$ (e.g., GaAs QDs \cite{KCPF}), there is strong spin mixing of the states at these points and
the SO term (\ref{eq:Hso}) leads to anticrossings of the corresponding levels \cite{BL}. Strong mixing of
spin-up and spin-down states and an anticrossing at $\omega_Z=\omega_-$ ($\omega_Z=3\omega_-$) are due to only DSO (RSO)
coupling. For $g<0$ (e.g., InAs QDs \cite{BKF}), the levels $E_{00\uparrow}$ and $E_{10\downarrow}$
($E_{00\uparrow}$ and $E_{30\downarrow}$) just cross each other at $\omega_Z=\omega_-$ ($\omega_Z=3\omega_-$), since
there is no coupling between the corresponding states. 

In the following we study spin relaxation induced by phonon-HH interactions only. The coupling between HHs and phonons with mode $\mathbf{q}\alpha$ ($\mathbf{q}$ is the
phonon wave vector, and the branch index $\alpha=L,T1,T2$ for one longitudinal and two transverse modes) is given by \cite{OO,KN}
\begin{eqnarray}
\nonumber U^{ph}_{\mathbf{q}\alpha}(\mathbf{r})&=&\sqrt{\frac{\hbar}{2\rho s_\alpha q V}}F(q_z)e^{i\mathbf{q_\parallel
r}}\\
\label{eq:U_ph} &&\times\left\{eA_{\mathbf{q}\alpha}+i\left[\left(a+\frac
b2\right)\mathbf{q\cdot d}^{\mathbf{q}\alpha}-\frac{3}{2}bq_zd_z^{\mathbf{q}\alpha}\right]\right\},
\end{eqnarray}
where $\rho$ is the crystal mass density, $s_\alpha$ is the sound velocity, $V$ is the volume of the QD,
$\mathbf{q}_\parallel=(q_x,q_y)$, $a$ and $b$ are the constants of the deformation potential,
$A_{\mathbf{q}\alpha}=\xi_i\xi_jd^{\mathbf{q}\alpha}_l\beta_{ijl}$, $\bm{\xi}=\mathbf{q}/q$,
$\mathbf{d}^{\mathbf{q}\alpha}$ is the phonon polarization vector, and $\beta_{ijl}$ is the piezotensor, which has
nonzero components only when all three indices $i,j,l$ are different: $\beta_{xyz}=\beta_{xzy}=\ldots=h_{14}/\varepsilon_S$
($\varepsilon_S$ is the static dielectric constant). 
For GaAs, $eh_{14}=1.2\times10^7\:$eV$/$cm, $\varepsilon_S=13.2$, and for InAs, $eh_{14}=3.38\times10^6\:$eV$/$cm, and
$\varepsilon_S=14.6$. In Eq.~(\ref{eq:U_ph}) we introduced the form-factor $F(q_z)$ which is determined by the spread of the HH wave function
 in the $z$-direction: $F(q_z)=\int dze^{iq_zz}|\psi_0(z)|^2$, where $\psi_0(z)$ is the ground state envelope wave
function of a HH along the $z$-direction. The form factor $F(q_z)$ equals unity for $|q_z|\ll d^{-1}$ and vanishes for
$|q_z|\gg d^{-1}$ \cite{GKL}.

We consider a single-particle QD, in which a HH can occupy one of the low-lying levels. As mentioned above, with
increasing $B$ some energy levels with the same spin orientation cross the upper Zeeman-split ground state level
and we should study the relaxation of an $n$-level system, the first $n-1$ levels have the same spin and the $n$-th
level has the opposite spin orientation. In the framework of Bloch -- Redfield theory \cite{Blum}, the Bloch equations of HH
spin motion for such a system in the interaction representation are given by
\begin{eqnarray}
\label{eq:Sz}
\langle \dot{S}_z\rangle&=&\left(S_T-\langle S_z\rangle\right)/T_1-R(t),\\
\langle \dot{S}_x\rangle&=&-\langle S_x\rangle/T_2,\ \langle \dot{S}_y\rangle=-\langle S_y\rangle/T_2,
\end{eqnarray}
where $R(t)=W_{n1}\rho_{nn}(t)+\sum_{i=1}^{n-1}W_{ni}\rho_{ii}(t)$, $\rho(t)$ is the density matrix, $W_{ij}$ is the
transition rate from state $j$ to state $i$, $S_T$ is a constant (which hase the value of $\langle S_z\rangle$ in thermodynamic equilibrium if $R(t)=0$), 
\begin{equation}
\label{eq:T1T2}
\frac{1}{T_1}=W_{n1}+\sum_{i=1}^{n-1}W_{in},\ \ \frac{1}{T_2}=\frac{1}{2T_1}+\frac12\sum_{i=2}^{n-1}W_{i1},
\end{equation}
where the pure dephasing (due to fluctuations along $z$ direction) is absent in the spin decoherence time $T_2$ since the spectral function is superohmic.
As can be seen from Eq.~(\ref{eq:Sz}), the spin motion has a complex dependence on the density matrix and, in the general
case, there are $n-1$ spin relaxation rates. However, in the case of low temperatures ($\hbar qs_\alpha\gg T$), when the
phonon absorption becomes strongly suppressed, solving the master equation, we find that $R(t)\approx0$,
therefore, there is only one spin relaxation time $T_1$. In this limit, the last sum in Eq.~(\ref{eq:T1T2}) is
negligible and the spin decoherence time saturates, i.e., $T_2=2T_1$.

For brevity, we present only the probability $W_{1n}$ of transition with phonon emission for the Zeeman-split ground
state which can be expressed as a sum of two terms due to RSO and DSO couplings, respectively:
$W_{1n}=W_{1n}^R+W_{1n}^D$, where
\begin{eqnarray}
\nonumber
W_{1n}^R&=&\frac{\alpha^2\hbar^3\omega_Z^7}{2^8\pi^2\rho\Omega^6}(N_{\omega_Z}+1)\left(\frac{\omega_+^3}{3\omega_%
++\omega_Z}-\frac{\omega_-^3}{3\omega_--\omega_Z}\right)^2\\
\label{eq:WR} &&\times\sum_\alpha s_\alpha^{-9}e^{-\omega_Z^2l^2/2s_\alpha^2}I^{(7)},\\
\nonumber W_{1n}^D&=&\frac{\beta^2m^2\hbar\omega_Z^3}{2^4\pi^2\rho\Omega^4}(N_{\omega_Z}+1)\sum_\alpha
s_\alpha^{-5}e^{-\omega_Z^2l^2/2s_\alpha^2}\phantom{\ \ \ \ \ \ \ \ \ \ \ \ \ \ \ \ \ }\\
\label{eq:WD}
&&\times\Bigg[f^2I^{(3)}+2fj\left(\frac{\omega_Zl}{2s_\alpha}\right)^2I^{(5)}+j^2\left(\frac{\omega_Zl}{2s_\alpha%
}\right)^4 I^{(7)}\Bigg].
\end{eqnarray}
Here $N_\omega=\left(e^{\hbar \omega/T}-1\right)^{-1}$, $l=\sqrt{\hbar/m\Omega}$,
$f=(\omega_-^2+\omega_+^2)[\omega_+/(\omega_++\omega_Z)-\omega_-/(\omega_--\omega_Z)]$,
$j=\omega_-\omega_+[\omega_+/(\omega_-+2\omega_++\omega_Z)-\omega_-/(2\omega_-+\omega_+-\omega_Z)]$,
\begin{eqnarray*}
I^{(k)}&=&\int_0^{2\pi}\!\!\!\!\!\!d\varphi\int_0^{\pi/2}\!\!\!\!\!\!d\vartheta\sin^k\vartheta 
F^2(\omega_Z\cos\vartheta/s_\alpha)e^{\omega_Z^2l^2\cos^2\vartheta/2s_\alpha^2}\\
&&\times\left\{\left(eA_{\mathbf{q}\alpha}\right)^2+\frac{\omega_Z^2}{s_\alpha^2}\left[\left(a+\frac
b2\right)\bm{\xi}\cdot\mathbf{d}^{\mathbf{q}\alpha}-\frac{3}{2}b\xi_zd_z^{\mathbf{q}\alpha}\right]^2\right\}.
\end{eqnarray*}
In the case of parabolic confinement along the growth direction of a QD, $I^{(k)}$ can be expressed in terms of 
error functions \cite{BL}. As mentioned above, for $g>0$, the SO term (\ref{eq:Hso}) leads to level anticrossings at
$\omega_Z=\omega_-$ (due to DSO) and $\omega_Z=3\omega_-$ (due to RSO). In this case, the denominators
$3\omega_--\omega_Z$ in Eq.~(\ref{eq:WR}) and $\omega_--\omega_Z$ in the expression for $f$ should be replaced by
$\mathrm{sgn}(3\omega_--\omega_Z)\sqrt{(3\omega_--\omega_Z)^2+(\Delta_R/\hbar)^2}$ and
$\mathrm{sgn}(\omega_--\omega_Z)\sqrt{(\omega_--\omega_Z)^2+(\Delta_D/\hbar)^2}$, respectively. Here
$\Delta_D=2\beta(ml)^3\omega_-(\omega_-^2+\omega_+^2)$ and $\Delta_R=2\sqrt{6}\alpha(ml\omega_-)^3$ are the level
splittings at the anticrossing points.

Note that there is no interplay between RSO and DSO couplings for HHs of a QD in perpendicular magnetic fields, as is true also for electrons \cite{GKL,BL}.

\begin{figure*}
\includegraphics[clip=true,width=17.8 cm]{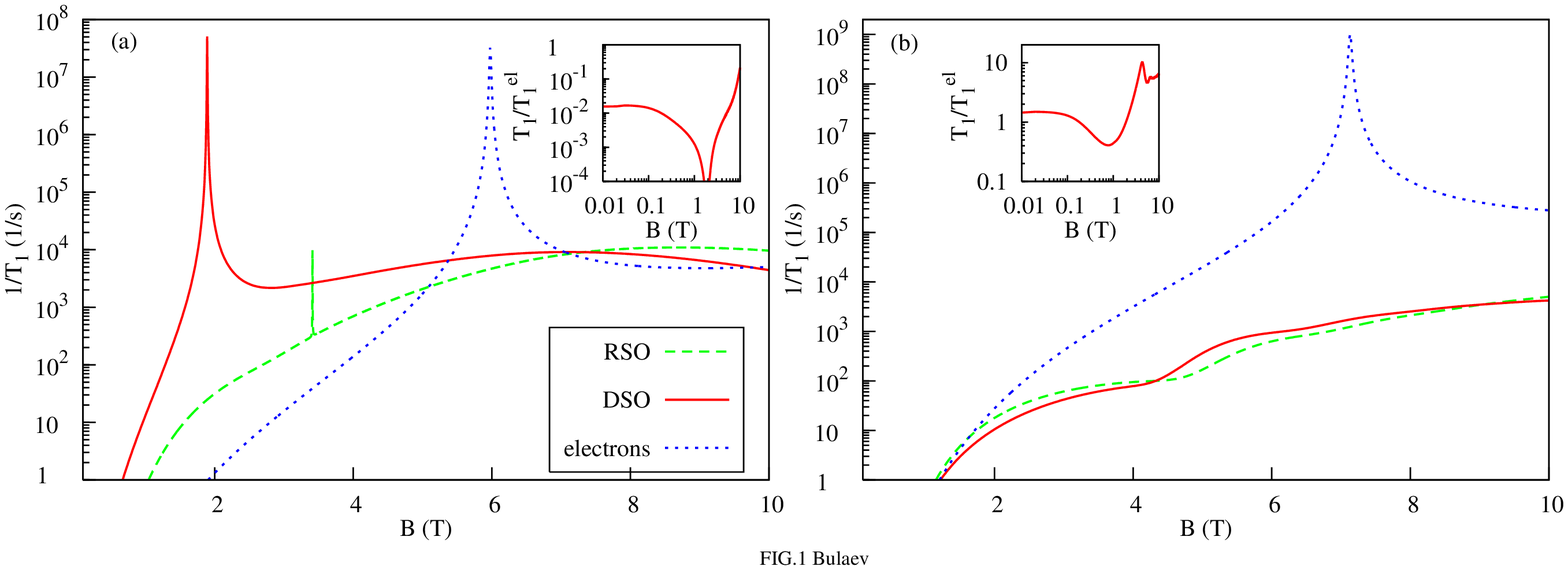}%
\caption{\label{fig:1} (color online). Spin relaxation rate $1/T_1$ of HHs (solid and dashed lines) and electrons (dotted line) in a GaAs (a) and an InAs (b) QD ($d=5\:$nm,
$l_0=\sqrt{\hbar/m\omega_0}=30\:$nm, and $T=0.1\:$K) due to DSO (solid line) and RSO (dashed line)
coupling ($\alpha=\beta$). For a GaAs QD (a), $g=2.5$ \cite{KCPF}, $\gamma/\hbar^3=28\:$eV$\:$\AA$^3$ \cite{MZM}, and $\Delta=40\:$meV \cite{SKG,WRK}, for
an InAs QD (b), $g=-2.2$ \cite{BKF}, $\gamma/\hbar^3=130\:$eV$\:$\AA$^3$ \cite{SRB}, and $\Delta=150\:$meV. We note that $T_2=2T_1$. Insets: Ratio between the 
HH ($T_1$) and electron ($T_{1}^{el}$) spin relaxation time due to DSO coupling.}
\end{figure*}
Figure~\ref{fig:1} shows the dependence of the spin relaxation rate $1/T_1$ for HHs in a GaAs (see
Fig.~\ref{fig:1}~(a)) and InAs (see Fig.~\ref{fig:1}~(b)) QD on a perpendicular magnetic field. The solid (dashed) curve corresponds
to the spin relaxation due to RSO (DSO) coupling. In the case of a positive HH $g$-factor (i.e., a GaAs QD), there are
peaks (see Fig.~\ref{fig:1}~(a)) in the relaxation rate curve at $\omega_{Z}=\omega_{-}$ ($\omega_{Z}=3\omega_{-}$),
which are caused by strong spin mixing at the anticrossing points due to DSO (RSO) coupling \cite{BL}. The half-width
of the first peak is $\Delta_{D}/\hbar$ and that of the second peak is $\Delta_{R}/\hbar$. Since, at anticrossing
points, $\omega_{+}\gg\omega_{-}$, the half-width of the peak, which is caused by the DSO coupling, is much larger than
that due to the RSO coupling (in our case, $\Delta_{R}=4\:$neV and $\Delta_{D}=1.5\:\mu$eV). In the case of
a negative HH $g$-factor (i.e., an InAs QD), there is no mixing of spin-up and spin-down states of the crossing levels,
therefore, the field dependence of the relaxation rate is monotonic (see Fig.~\ref{fig:1}~(b)). 

From Eqs.~(\ref{eq:WR}) and (\ref{eq:WD}) it can be shown that at low temperatures and at low magnetic fields
($B<0.5\:$T) the relaxation rate due to RSO (DSO) coupling $\sim B^{9}$ ($\sim B^{5}$). Therefore, the field dependence of
$T_{1}$ due to DSO coupling of HHs in a QD is the same as for electrons \cite{KN,K2004,BL}, but that due to
RSO coupling is quite different. This qualitative difference in the field dependence of the spin relaxation can serve to provide information about the leading SO interaction term at low $B$.

Let us consider the quantitative difference between the spin relaxation of electrons and HHs. For simplicity, we consider DSO coupling
only. It can be shown that at low $B$ ($B\le0.1\:$T) the ratio of the HH spin relaxation time
 and that for electrons $T_{1}^{el}$ is given by
\begin{equation}
\label{eq:Teh}
\frac{T_{1}}{T_{1}^{el}}\approx\frac{16}{9}\left(\frac{g_{el}}{g}\right)^4\left(\frac{m_{el}}{m}\right)^4\left(\frac{l_0}{d}\right)^4\eta^2,
\end{equation}
where $g_{el}$ and $m_{el}$ are the electron $g$-factor and effective mass, respectively. Usually the $g$-factor and
the effective mass of an electron are less than those of a HH and the spin relaxation time $T_1^{el}$ for electrons is much longer than for HHs. However, for strongly 2D QDs ($l_0\gg d$), the spin relaxation time
for HHs can be comparable to, or even longer than, that for electrons. For the  GaAs QD considered here ($\eta=0.18$), $T_{1}$ for
HHs is comparable to that for electrons \cite{GKL,BL} (see inset Fig.~\ref{fig:1}~(a)). Now consider the InAs QD, for which $g_{el}$ is larger than
for the GaAs QD and the energy gap is narrow ($\eta=0.48$). We find that the spin relaxation time for HHs is longer than for electrons (see inset in Fig.~\ref{fig:1}~(b)).

Since $\alpha$ and $\beta\sim d^{2}$, the spin relaxation rate increases with increasing confinement along the growth direction of a QD and, as can be seen from Eqs.~(\ref{eq:WR}) and (\ref{eq:WD}), with decreasing lateral confinement (with a decrease in the confinement frequency $\omega_{0}$).
As follows from Eq.~(\ref{eq:U_ph}), the HH spin relaxes primarily due to piezoelectric phonons at low $B$ and due to deformational acoustic phonons at $B>1\:$T.

In conclusion, we have shown that due to the different symmetries of the RSO and DSO terms in momentum space, these terms lead to different behavior of the spin relaxation: at low magnetic fields $T_1\sim B^{-9}$ in the case of RSO coupling and $T_1\sim B^{-5}$ in the case of DSO coupling. The field dependence of the spin relaxation rate is monotonic for a system with a negative $g$-factor (i.e., HHs in InAs QDs), whereas for $g>0$ (i.e., HHs in GaAs QDs) the relaxation rate has peaks corresponding to level anticrossings and the associated enhanced mixing of spin-up and spin-down states at the anticrossings.

\begin{acknowledgments} We thank V.N.~Golovach, W.A.~Coish, J.~Lehmann, and A.~Khaetskii for useful discussions. We acknowledge
support from the Swiss NSF, NCCR Basel, EU RTN ``Spintronics'', U.S. DARPA, ARO, and ONR.
\end{acknowledgments}


\begin{thebibliography}{33}
\bibitem{ALS}
\emph{Semiconductor Spintronics and Quantum Computing}, edited by
 D.~D.~Awschalom, D.~Loss, and N.~Samarth 
 (Springer, New York, 2002).

\bibitem{LV}
 D.~Loss and D.~P.~DiVincenzo,
  Phys. Rev. A \textbf{57}, 120 (1998).

\bibitem{RBMCh}
 D.~Rugar,
  R.~Budakian,
  H.~J.~Mamin, and
  B.~W.~Chui,
  Nature \textbf{430},
  329 (2004).

\bibitem{E2004}
 J.~M.~Elzerman \textit{et al.}, Nature
 \textbf{430}, 431 (2004).

\bibitem{K2004}
 M.~Kroutvar \textit{et al.}, Nature
 \textbf{432}, 81 (2004).

\bibitem{KN}
 A.~V.~Khaetskii and Y.~V.~Nazarov, Phys. Rev. B
 \textbf{64}, 125316
 (2001).

\bibitem{GKL}
 V.~N.~Golovach, A.~Khaetskii, and D.~Loss,
  Phys. Rev. Lett. \textbf{93},
  016601 (2004).

\bibitem{BL}
 D.~V.~Bulaev and D.~Loss,
  cond-mat/0409614.

\bibitem{FA}
 V.~I.~Fal'ko, B.~L.~Altshuler, and O. Tsyplyatev,
 cond-mat/0501046.

\bibitem{EN}
 S.~I.~Erlingsson and Y.~V.~Nazarov, Phys. Rev. B \textbf{66}, 155327 (2002).

\bibitem{KLG}
 A.~V.~Khaetskii,
  D.~Loss, and L.~Glazman,
  Phys. Rev. Lett. \textbf{88},
  186802 (2002).

\bibitem{CL}
 W.~A.~Coish and D.~Loss,
  Phys. Rev. B \textbf{70},
  195340 (2004).

\bibitem{SKG}
 P.~Schneider \textit{et al.}, J. Appl. Phys
 \textbf{96}, 420 (2004).

\bibitem{SFS}
 G.~Sun,
  L.~Friedman, and
  R.~A.~Soref,
  Phys. Rev. B \textbf{62},
  8114 (2000).

\bibitem{Dresselhaus}
 G.~Dresselhaus,
  Phys. Rev. \textbf{100},
  580 (1955).

\bibitem{BR}
 Y.~Bychkov and
  E.~I.~Rashba,
  J. Phys. C \textbf{17},
  6039 (1984).

\bibitem{ER}
 Al.~L.~Efros and M.~Rosen,
 Phys. Rev. B \textbf{58} , 7120 (1998).

\bibitem{WRK}
 L.~M.~Woods, T.~L.~Reinecke, and R.~Kotlyar,
  Phys. Rev. B \textbf{69},
  125330 (2004).

\bibitem{LCW}
 C.~L{\"u}, J.~L.~Cheng, and M.~W.~Wu,
  Phys. Rev. B \textbf{71},
  075308 (2005).

\bibitem{OO}
 \emph{Optical Orientation}, edited by B.~P.~Zakharchenya and F.~Meier
 (North-Holland, Amsterdam, 1984).

\bibitem{LK}
 J.~M.~Luttinger and W.~Kohn,
  Phys. Rev. \textbf{97},
  869 (1955).

\bibitem{Luttinger}
 J.~M.~Luttinger,
  Phys. Rev. \textbf{102},
  1030 (1956).

\bibitem{Winkler1}
 R.~Winkler, Phys. Rev. B \textbf{62},
  4245 (2000).

\bibitem{Winkler2}
 R.~Winkler,  H.~Noh,  E.~Tutuc, and  M.~Shayegan,
  Phys. Rev. B \textbf{65},
  155303 (2002).

\bibitem{KCPF}
 H.~W.~van Kesteren,  E.~C.~Cosman, W.~A.~J.~A.~van~der Poel, and C.~T.~Foxon, Phys. Rev. B
 \textbf{41}, 5283 (1990).

\bibitem{GMSh}
 N.~G.~Galkin, V.~A.~Margulis, and A.~V.~Shorokhov, Phys. Rev. B
 \textbf{69}, 113312
 (2004).

\bibitem{GBG}
 S.~D.~Ganichev \textit{et al.},
  Phys. Rev. Lett. \textbf{92}, 256601 (2004).

\bibitem{BKF}
 M.~Bayer \textit{et al.},
  Phys. Rev. Lett. \textbf{82},
  1748 (1999).

\bibitem{Blum}
 K.~Blum,
 \emph{Density Matrix Theory and Applications}
 (Plenum Press, New York, 1996).

\bibitem{MZM}
 J.~Miller \textit{et al.},
  Phys. Rev. Lett. \textbf{90},
  076807 (2003).

\bibitem{SRB}
 E.~A.~de~Andrada~e Silva,  G.~C.~La~Rocca, and F.~Bassani,
  Phys. Rev. B \textbf{50},
  8523 (1994).
\end{thebibliography}

\end{document}